\documentclass[aps,pre,twocolumn,superscriptaddress]{revtex4-1}
\usepackage[utf8]{inputenc}

\usepackage{graphicx}
\usepackage{epstopdf}
\usepackage{bm}
\usepackage{amsmath}
\usepackage{verbatim}
\usepackage{amssymb}
\usepackage{listings}
\usepackage{color}
\definecolor{darkblue}{rgb}{0,0,0.6}
\usepackage{here}
\usepackage[colorlinks,linkcolor=darkblue,citecolor=darkblue,urlcolor=darkblue]{hyperref}
\DeclareRobustCommand{\average}[1]{\left\langle #1 \right\rangle}
\renewcommand\vec[1]{\boldsymbol{#1}}

\newcommand\Fig[1]{Fig.~\ref{#1}}
\newcommand\Sec[1]{Sec.~\ref{#1}}
\newcommand\App[1]{Appendix~\ref{#1}}

\renewcommand\vec[1]{\boldsymbol{#1}}
\newcommand\eq[1]{Eq.~(\ref{#1})}

\newcommand\bigpar[1]{\left( #1 \right)}
\begin{document}

\title{Stationary Bootstrap: \\A Refined Error Estimation for Equilibrium Time Series}
\author{Yoshihiko Nishikawa}
\affiliation{Laboratoire Charles Coulomb (L2C), Universit\'e de Montpellier, CNRS, 34095 Montpellier, France}
\affiliation{Graduate School of Information Sciences, Tohoku University, Sendai 980-8579, Japan}

\author{Jun Takahashi}
\affiliation{
Institute of Physics, Chinese Academy of Sciences, Beijing 100190, China}
\affiliation{
Department of Physics, Boston University, 590 Commonwealth Avenue, Boston, Massachusetts 02215, USA}
\affiliation{
Center for Quantum Information and Control, 
University of New Mexico, Albuquerque, New Mexico 87131, USA}

\author{Takashi Takahashi}
\affiliation{
    Institute for Physics of Intelligence 
  \\ Department of Physics Graduate School of Science
  \\ The University of Tokyo 7-3-1 Hongo, Bunkyo-ku, Tokyo 113-0033, Japan
}
\date{\today}

\begin{abstract}
In Markov-chain Monte Carlo simulations, estimating statistical errors or confidence intervals of numerically obtained values is an essential task. In this paper, we review several methods for error estimation, such as simple empirical estimation with multiple independent runs, the blocking method, and the stationary bootstrap method. We then study their performance when applied to an actual Monte-Carlo time series. We find that the stationary bootstrap method gives a reasonable and stable estimation for any quantity using only one single time series. In contrast, the simple estimation with few independent runs can be demonstratively erroneous. We further discuss the potential use of the stationary bootstrap method in numerical simulations.
\end{abstract}

\maketitle

\section{Introduction}
The Monte Carlo method is a general, versatile tool to solve an extremely wide range of problems in science, including condensed-matter and high-energy physics \cite{Landau2009,Berg2004,Krauth2006,Tanabashi2018}, biology \cite{Manly2018}, and machine learning \cite{Andrieu2003}. 
The progress of modern computers allows us to perform Monte Carlo calculation of statistical problems with millions of degrees of freedom within a reasonable amount of time and estimate various quantities with great precision. 
However, because Monte Carlo method estimates quantities from a finite number of samples, the results inevitably have statistical uncertainties and fluctuate from simulation to simulation. 
Accurately estimating statistical errors, or the confidence intervals, is thus as essential as computing the quantities themselves in interest.
Note that the confidence interval is the fluctuation regarding the sampling distribution, not to be confused with the physically relevant distribution of the observable (see \Sec{sec:statistics} for more discussions). Assessing the fluctuation of the sampling distribution is what is needed for estimating the statistical errors.

When dealing with samples that are independent and identically distributed (i.i.d.), we know the sampling distribution will converge to a Gaussian distribution thanks to the central limit theorem. This allows us to use the standard error of many samples to construct the $68\%$ confidence interval quite reliably. More sophisticated resampling methods, including the jackknife and the bootstrap methods \cite{Miller1974,Efron1979,Efron1994,Wasserman2004,Wasserman2006}, have been proposed and rigorous studies have proven asymptotic exactness in the limit of infinite number of samples for some setups. 
Those methods virtually enhance the number of nearly-independent data, enabling more accurate estimates of the statistical error. With the advances of these resampling methods, the statistical analysis for i.i.d. samples have now been established well.

For most of the statistical problems in computational physics, we use the Markov-chain Monte Carlo (MCMC) method to produce samples. Such samples often have correlation in the time direction. This correlation complicates their statistical analysis and the statistical technique for i.i.d. samples mentioned above is unavailable in such cases.
The most straight-forward and commonly used way to deal with the statistical correlations is to have several totally independent runs and then calculate the standard error among them as the estimated statistical error.
Heuristic methods such as the blocking method \cite{Flyvbjerg1989,Newman1999,Krauth2006,Jonsson2018}
has also been used by physicists.

On the other hand, in the field of modern computational statistics, various frameworks for estimating the statistics of time-correlated samples, with no assumption on its sampling distribution have been proposed \cite{Carlstein1986,Kunsch1989,Liu1992,Politis1992,Politis1994}. Despite these advances, the extensive knowledge in statistics has not been fully imported to the field of computational physics.
We aim to shed light on this topic by quantitatively analyzing how the modern method from statistics compares to the traditional approach used commonly in computational physics.

\begin{figure}[bthp]
\includegraphics[width=\linewidth]{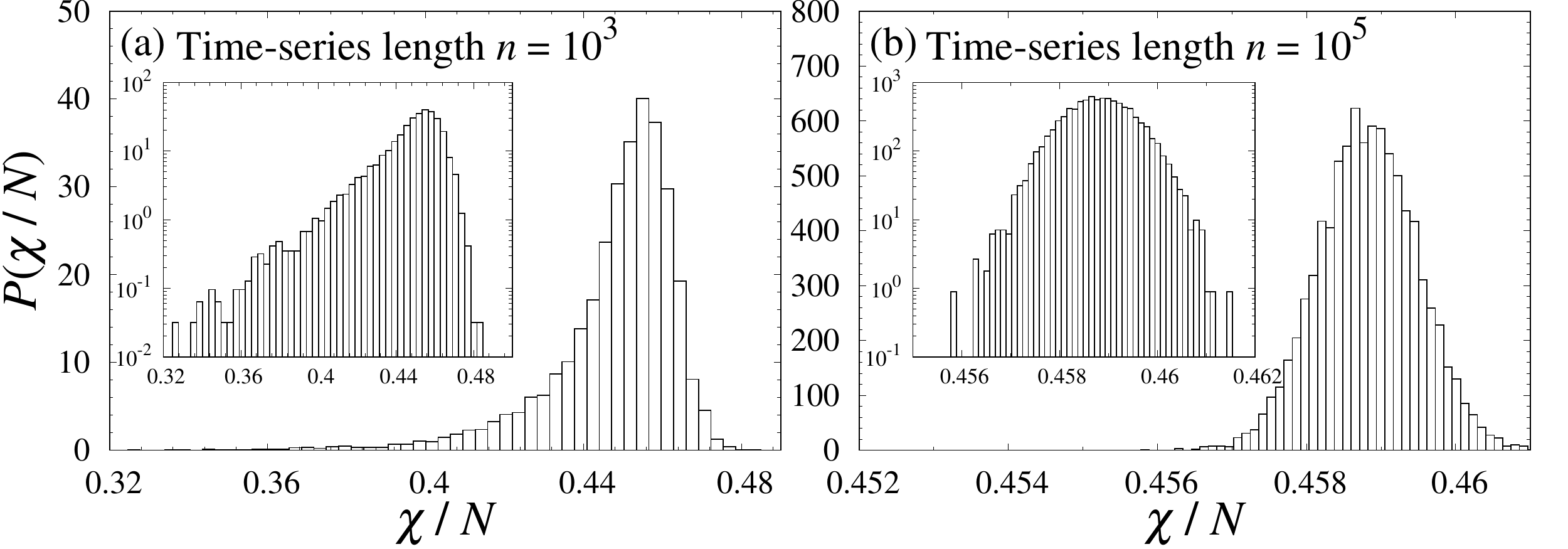}%
\caption{The sampling distribution of the susceptibility $\chi/N$ of actual Monte Carlo data of the two-dimensional Ising model (see \Sec{sec:num} for the details of our numerical simulation), with the time-series length $n=10^3$ (a) and $10^5$ (b), respectively. Each inset shows a semi-log plot of each panel.}
\label{fig:NonGaussianIsBad}
\end{figure}
In this paper, we focus on the error estimation of stationary time-series data with a finite correlation time generated by MCMC methods.
Applying several statistical methods to numerically obtained samples, we illustrate how each method estimates the confidence interval for various number of samples, and contrast their performances in a realistic setup. 
In order to be specific, here we define the statistical errors as confidence intervals of the computed physical quantities, which indicate high probability of the {\it true} value existing within the interval. This ``high probability'' is usually set to $\mathrm{erf}(1/\sqrt{2})\approx 0.68$ corresponding to the value of the standard error of a Gaussian distribution \footnote{In experimental fields, usually a 95\% bar is set.}. The standard error indeed gives a reliable estimation of the $68\%$ confidence interval when the number of samples is large enough and the sampling distribution is well-approximated by a Gaussian distribution. When the sampling distribution is strongly non-Gaussian (see \Fig{fig:NonGaussianIsBad}), however, the straight-forwardly calculated standard error will tend to underestimate (be overconfident in) the confidence interval.

Perhaps the simplest method of error estimation used frequently in the MCMC community is calculating the standard error from several independent runs \cite{Krauth2006,Wang2015,Nishikawa2016,Anderson2017}.
Although this method is straightforward and easy to implement, it obviously demands more computational resources with a multiplicative factor of the number of independent runs. In some cases where computational physicists attempt to reach the state-of-the-art limit with very large system sizes etc., it is not the most efficient way. 
Importantly, when the number of independent runs is small, this method would underestimate the statistical error.

Another well-used technique is the ``blocking'' (or ``binning'') method \cite{Flyvbjerg1989,Newman1999,Krauth2006,Jonsson2018} where the time series is divided into $K\gg 1$ ``blocks'' of certain fixed time periods. Each block can yield a rather noisy estimate of the desired quantity, and the statistical error is calculated from the standard error among those $K$ values. By combining two consecutive blocks as one data, we can have another estimate of the error from the $K/2$ combined ``double blocks''. 
We can go on with more combining steps, which could be interpreted as renormalization in the time direction \cite{Flyvbjerg1989}. The expectation is that the estimated error will converge to the ``correct'' value as long as the renormalization steps $k$ are not too large (ending up giving only $K/2^k\rightarrow O(1)$ data). However, when the time-series length is finite, it is hard to systematically determine whether the number of blocks $K$ and renormalization steps $k$ are appropriate. While this method can give reasonably good estimation of error in practice, it is difficult in general to find if the estimation is precise since such ``convergent renormalization steps'' $k_\text{opt}$ may not even exist for a particular finite data set.

The paper is organized as follows. In \Sec{sec:statistics}, we summarize some basic concepts in statistics and comment on the meaning of statistical errors in general. We also remind the difficulty of evaluating statistical errors in numerical simulations of physical systems.
In \Sec{sec:bootstrap}, we review the bootstrap method proposed by B.~Efron \cite{Efron1979,Efron1994}. We then introduce the stationary bootstrap method in \Sec{sec:stbootstrap} and its practical algorithm. In \Sec{sec:num}, we detail our numerical procedures for producing time series and estimating the errors. We then discuss the performance of the stationary bootstrap method, together with comparison to the error estimation using independent runs and the blocking method in \Sec{sec:results}. We conclude in the last section \Sec{sec:summary}, with discussing practical guidelines of the stationary bootstrap method for practical situations.

\section{Error estimation in sampling methods}
\label{sec:statistics}
In MCMC, a time series $x_1, x_2, \cdots, x_n$ drawn from a distribution $q$ is produced, where $x_t$ is a physical quantity of interest, such as the magnetization of the Ising model or the microscopic configuration of the physical system itself, at time step $t$. 

Suppose we are interested in estimating a physical quantity $A$. The physical quantity $A$ is evaluated from the time series as 
\begin{equation}
    \bar{A}_n = \bar{A}_n(x_1, x_2, \cdots, x_n).
    \label{eq:estimated physical quantity}
\end{equation} 
We expect that $\bar{A}_n$ converges to $A$ in the large sample limit $n\to\infty$.
However, this estimator $\bar{A}_n$ depends on the realization of the time series, e.g. the seed of a random number generator etc., which motivates us to evaluate its statistical error. 

Again, we stress that the estimation of the confidence interval is about the sampling distribution
\begin{equation}
    P_n(a) = \average{\delta(a - \bar A_n)},
    \label{eq:sampling dist}
\end{equation}
where $\delta$ is Dirac's delta function. 
This distribution is different from the distribution of a physical quantity, such as the magnetization and the energy density.
Here, we denote by $\average{\cdots}$ an average over the realizations of time series,
\begin{equation}
    \average{\cdots} = \int (\cdots) q(x_1, x_2, \cdots, x_n)dx_1dx_2\cdots dx_n.
\end{equation}
The sampling distribution clearly depends on $n$ and it converges to a delta function at $n \to \infty$.
On the other hand, the physical distribution is independent of $n$ and has a non-zero variance in general.

As an example, we consider the case where $x_i$ $(i = 1, \cdots n)$ is a physical quantity that is defined for each microscopic configuration of the system (i.e. sample), such as the magnetization and the energy density. 
Then the distribution of that quantity could be estimated as
\begin{equation}
    P(x) = \average{\frac1n\sum_i \delta(x - x_i)}.
\end{equation}
This converges to a physical distribution (e.g. the Boltzmann distribution), and is thus a physically measurable quantity.
It has a non-zero variance even in the limit $n\to\infty$ and its fluctuation has an inherent physical meaning; e.g. the specific heat when $x_i$ is the energy density.
On the other hand, the {\it sampling distribution} of the sample mean $\sum_{i}x_i / n$,
\begin{equation}
    P_n(a) = \average{\delta\left(a - \frac{1}{n}\sum_i x_i\right)},
\end{equation}
which simply measures the fluctuation among various estimation values. 
Even when $P(x)$ is strongly non-Gaussian, e.g. asymmetric or multimodal, the sampling distribution $P_n(a)$ converges to a Gaussian distribution in the limit $n\to\infty$ if the variance of $P(x)$ is finite, as the central limit theorem states. These two distributions must not be confused.

For measures of the statistical error, we would primarily focus on the standard error $\sigma_{A, n}$, i.e. the variance of the sampling distribution $P_n(a)$, and the $68\%$ confidence interval $I_{A, n}$ defined as follows: 
\begin{align}
    \sigma_{A,n}^2 &= \average{\left(\bar{A}_n\right)^2} - \average{\bar{A}_n}^2,
    \label{eq:a_var}
    \\
    I_{A, n} &= [L_{A,n}, U_{A, n}],
    \nonumber \\
    &{\rm s.t.~}{\rm Prob}\left[
        \lim_{n\to\infty}\bar{A}_n\in I_{A,n}
    \right] = 0.68.
    \label{eq:a_conf}
\end{align}
Estimating lower and upper limits of the interval ($L_{A,n}$ and $U_{A, n}$) from data yields the confidence interval.
When $\bar A_n$ fluctuates around $\lim_{n\to\infty}\bar{A}_n$ with a Gaussian noise, $[\bar{A}_n - \sigma_{A,n}, \bar{A}_n + \sigma_{A,n}]$ converges to the $68\%$ confidence interval in the $n\rightarrow\infty$ limit. 
When $n$ is finite, however, we often find a non-Gaussian fluctuation and the standard error does not represent the $68\%$ confidence interval.

In general, finding the analytic form of the sampling distribution for a generic physical quantity is unfeasible. Instead, one could directly estimate the sampling distribution by running as many MCMC runs as possible, although it is computationally very demanding. Thus, it would be nice if we can estimate the sampling distribution \eq{eq:sampling dist} from a single set of samples with no knowledge on its analytic form. This kind of estimation problem is known as the {\it nonparametric} inference \footnote{The name comes from the fact that when we have knowledge of the distribution form, then the problem essentially boils down to estimating parameters for such distributions.} in the literature of statistics \cite{Wasserman2006}. For i.i.d. samples, Efron's bootstrap method \cite{Efron1979} and its improved versions \cite{Efron1994} are known to work quite well. When the samples have correlations, however, Efron's bootstrap does not work and some adaptation taking the correlation into account is necessary. For the case of time correlations, we will demonstrate one of the nonparametric bootstrap methods indeed gives the statistical properties accurately in a typical situation.

We must also mention that in general, the convergence of the time series to the stationary state becomes a serious issue with MCMC. However, since it is more of a matter of the MCMC algorithms themselves rather than the error estimation technique, in this paper we will assume that the MCMC has been well-converged after the initial equilibration.

\section{The Bootstrap method}
\label{sec:bootstrap}
To resolve the aforementioned difficulties for evaluating statistical errors from time-correlated data, statisticians have developed various nonparametric methods. Among the many methods \cite{Carlstein1986,Kunsch1989,Liu1992,Politis1992}, the {\it stationary} bootstrap \cite{Politis1994}, based upon Efron's original bootstrap method for i.i.d. samples \cite{Efron1979}, is one of the most versatile methods. 

To explain the stationary bootstrap method, we first describe Efron's bootstrap method developed for analyzing i.i.d. samples \cite{Efron1979}, following the standard textbook \cite{Efron1994} in the next subsection.

\subsection{Efron's bootstrap}
\label{sec:iidboostrap}
To describe Efron's bootstrap method, let us assume that the samples have no correlation in this subsection. In this case, the joint distribution $q(x_1, x_2, \cdots, x_n)$ of the data points $x_1,x_2,\dots,x_n$ can be factorized with a probability distribution $q_0$ as follows:
\begin{equation}
    q(x_1,x_2,\dots, x_n) = \prod_{i=1}^n q_0(x_i).
\end{equation}
If we could generate new samples from $q_0$ at a low computational cost, then we can directly evaluate the sampling distribution in \eq{eq:sampling dist} by generating as many samples as needed. However, generating new samples is often computationally demanding, and thus, we will approximate the distribution $q_0$ by some computationally tractable distribution.

The key observation here is that when the samples are independent, the empirical distribution $q_{\rm emp}$ constructed by many samples approximates $q_0$ quite precisely:
\begin{equation}
    q_{\rm emp}(x):=\frac{1}{n}\sum_{i=1}^n\delta(x - x_i) \simeq q_0(x), \quad n\gg1,
    \label{eq:empirical dist}
\end{equation}
and that $q_{\rm emp}$ is unbiased in the following sense:
\begin{equation}
    \langle q_{\rm emp}(x)\rangle 
    = q_0(x), \quad \forall n=1,2,\dots.
\end{equation}
Furthermore, sampling from the empirical distribution is computationally easy because we can generate a new set of samples just by sampling from the already obtained data $x_1, x_2, \cdots, x_n$ with replacement.
These observations suggest that the empirical distribution is a good approximation of $q_0$ that can be constructed from a single set of samples.
Note that the above observations do not apply for the case of any form of correlated samples because the empirical distribution cannot serve as an approximation of the true generation process of such samples even at $n\to\infty$.

In practice, the bootstrap method numerically computes the average in \eq{eq:sampling dist} by resampling pseudo data set from the empirical distribution \eq{eq:empirical dist} as follows:
\begin{enumerate}
    
    \item Construct the empirical distribution $q_{\rm emp}$ from the data as in \eq{eq:empirical dist}.
    
    \item For $b=1,2,\cdots, B$, repeat the following. Here $B$ is a parameter that specifies how many pseudo data sets are to be sampled.
    \begin{enumerate}
        \item Sample $n$ new data point $x_1^{*(b)}, x_2^{*(b)}, \cdots, x_n^{*(b)}$ from the empirical distribution $q_{\rm emp}$. This pseudo data set is called the bootstrap sample.
        \item Calculate the estimated value of the physical quantity $A$ as in \eq{eq:estimated physical quantity} for this pseudo data set $\bar{A}_n^{*(b)} = A(x_1^{*(b)}, x_2^{*(b)}, \cdots, x_n^{*(b)})$.
        
    \end{enumerate}
    
    \item Approximate the sampling distribution in \eq{eq:sampling dist} as 
    \begin{equation}
        P_n(a) \simeq \tilde{P}_{n,B}(a) := \frac{1}{B}\sum_{b=1}^B \delta(a - \bar{A}_n^{*(b)}).
        \label{eq:bootstrap dist for a}
    \end{equation}

\end{enumerate}
The confidence interval for $\bar{A}_n$ can be calculated from the approximated distribution obtained in \eq{eq:bootstrap dist for a}. For example, the most naive approximation for the $68\%$ confidence interval, which is known as the bootstrap percentile method \cite{Efron1994,DiCiccio1996}, yields $[L^*, U^*]$, where $L^*$ and $U^*$ represent the $16$ and the $84\%$ quantiles of $\tilde{P}_n(a)$, respectively. 
Despite its simplicity, this estimation often has a first-order accuracy \cite{Hall1988}, i.e. $|0.68 - {\rm Prob}[\lim_{n\to\infty}\bar{A}_n \in [L^*, U^*]| = O(n^{-1/2})$ for sufficiently large $n$. Thus, the bootstrap method gives a reasonable estimate of the confidence interval for any physical quantity without knowing its analytic form from a single time-series data when the data satisfies the i.i.d. assumption.

\subsection{Stationary bootstrap}
\label{sec:stbootstrap}

Whereas Efron's original bootstrap method \cite{Efron1979,Efron1994} provides systematic, nonparametric estimation for statistical properties, its applicability is limited to the i.i.d. case. The bootstrap method has been argued to give a reasonable estimation even when applied to correlated samples \cite{Newman1999},
however it actually turns out that the simple bootstrap method underestimates the statistical error by a few orders of magnitude, i.e. it fails to correctly estimate the sampling distribution even in the limit $n \to\infty$ (see \App{app:original_bootstrap} for further discussions). Thus we should not use the i.i.d. bootstrap method for error estimation of correlated samples.

For dependent samples with time correlations, several variants of the bootstrap method were proposed \cite{Carlstein1986,Kunsch1989,Liu1992,Politis1992,Politis1994}. Those methods build bootstrap samples by taking blocks of contiguous samples from a time series and piecing them together. The only parameter for those algorithms is the (average) size of the blocks.
With an appropriate size of blocks, each block can be considered as an independent short time series and a pseudo time series made by piecing these blocks together have similar statistical properties to the original time series in an asymptotic setting.
In some limiting cases, the literature rigorously proves that the methods actually give the exact statistical properties.
Here, we review the stationary bootstrap method \cite{Politis1994} that respects the stationarity of the time series, unlike other variants of the bootstrap method.

The stationary bootstrap assumes that a time series $x_1, x_2, \cdots, x_n$ is strictly stationary and weakly dependent, i.e. the time series is time-translation invariant and its time correlation asymptotically decays to $0$ with time. These conditions are typically met for a well-relaxed MCMC data. With these assumptions, we impose a `periodic boundary condition' for the time series, $x_{n+k}=x_k$, and build bootstrap samples. The practical procedures of the stationary bootstrap method is the following:

\begin{enumerate}
\item[0.] Set a parameter $p\in(0,1)$. This parameter controls the typical size of the blocks sampled in the following step.
\item Build $B$ bootstrap pseudo time series
\begin{equation*}
    \vec x^{*(b)} = (x^{*(b)}_1, x^{*(b)}_2, \cdots, x^{*(b)}_n), \ b = 1, 2,\cdots, B,
\end{equation*}
as follows:
\begin{enumerate}
\item Choose $t_1 \in [1, n]$ at random and let $x^*_1 = x_{t_1}$.
\item With probability $1-p$, let $t_{i+1} = t_i + 1$, and with probability $p$, choose $t_{i+1} \in [1, n]$ at random.
Then let $x^*_{i+1} = x_{t_{i+1}}$. Regard $x_{n+k}$ as $x_k$.
\label{item:b}
\item Repeat (\ref{item:b}) until $i = n-1$.
\end{enumerate}
\item For each bootstrap pseudo time series $\vec x^{*(b)}$, calculate any quantity of interest $\overline A^{*(b)}_n = A(\vec x^{*(b)})$, 
e.g. $\overline{A}_n^{*(b)}(\vec x^{*(b)}) = \sum x_i^{*(b)} / n$ if $A$ is the sample mean.
\item Approximate the sampling distribution of $A$
\begin{equation}
    P_n(a) \simeq P^*_{n,B}(a) := \frac1B \sum_{b=1}^B \delta(a - \overline A^{*(b)}_n).
\end{equation}
This enables calculation of any statistical quantity of the distribution, the most important values being the mean value as the estimator of the quantity and the standard error as the confidence interval. 
\end{enumerate}
In this algorithm, the number $B$ of bootstrap samples should be large enough to have the estimation almost independent of $B$. We carefully checked the $B$ dependence of the results, and set $B = 10^3$ throughout this paper.

The probability $p$, which determines the typical length of the blocks of successively chosen samples, controls the estimated value of the confidence interval. 
Choosing an appropriate value for $p$ is essential in the stationary bootstrap to handle the effect of correlation among the original time-series data.
It was rigorously shown that having the conditions $p \to 0$ and $np \to \infty$ results in exact estimation in the limit $B,n \to \infty$ \cite{Politis1994}, but the optimal value for a given time series with a finite length is \textit{a priori }unknown. In practice, we choose $p$ to minimize the mean-squared error of the coefficient of the estimated standard error of the simple mean, $n\sigma^2_{n, \text{SB}}$, using the stationary bootstrap method, \cite{Lahiri1999,Politis2004,Nordman2009,Patton2009},
\begin{align}
    \text{MSE}(n\sigma_{n,\text{SB}}^2) 
    &:= \left(\left\langle n\sigma_{n,\text{SB}}^2\right\rangle - n\sigma^2_n\right)^2 \nonumber \\
    &\quad  + \left\langle n\sigma_{n,\text{SB}}^2 - \left\langle n\sigma_{n,\text{SB}}^2\right \rangle\right\rangle^2 \\
    &= G^2 p^2 + \frac{D}{pn} + o(p^2) + o((pn)^{-1}),
    \label{eq:mse}
\end{align}
yielding 
\begin{equation}
    p_\text{opt} = \bigpar{\frac{D}{2nG^2}}^{1/3},
    \label{eq:popt}
\end{equation}
where 
\begin{align}
    &\sigma^2_n = \left\langle \left(\sum x_i / n\right)^2 \right\rangle - \left\langle \sum x_i / n\right\rangle^2,\\
    &G = 2\sum_{t=0}^\infty |t| C(t), \label{eq:G} \\
    &D = 2\bigpar{C(0) + 2\sum_{t=1}^\infty C(t)}^2 \label{eq:D},
\end{align}
and the unnormalized autocorrelation function
\begin{equation}
    C(t) = \average{\bigpar{x_0 - \average{x_i}}\bigpar{x_t - \average{x_i}}}.
    \label{eq:unautocorr}
\end{equation}
\eq{eq:popt} automatically satisfies the rigorous condition for the correct estimation, 
i.e. $p_\text{opt} \sim n^{-1/3} \to 0$ and $np \to \infty$ at $n\gg1$, if $D/G^2$ is independent of $n$. Using these formulae, fortunately, the optimal value of $p$ can also be estimated from the time series we wish to analyze \cite{Politis2004,Patton2009} (see \App{app:popt} for a detailed algorithm to find $p_\text{opt}$). Note that, when samples are i.i.d., $p_\text{opt} = 1$, meaning that the stationary bootstrap becomes equivalent to the original bootstrap method in this case.

The precise way we create these bootstrap samples are actually crucial.
For example, one may wonder what happens if we take a single block of contiguous samples  
whose length is distributed according to the geometric distribution with mean $\hat{p}^{-1}_{\rm opt}$ as the bootstrap sample $\vec{x}^{*(b)}$.
In this case, the length of the bootstrap time series fluctuates with variance $\hat{p}^{-2}_{\rm opt} -\hat{p}^{-1}_{\rm opt} \simeq \hat{p}^{-2}_{\rm opt}~(\hat{p}_{\rm opt}\ll 1)$, 
and because of that fluctuation the procedure fails to give the correct statistical properties.

Another reasonable attempt that actually works is to randomly take blocks with a fixed length $\hat p^{-1}_\text{opt}$ of contiguous samples out of the original time series, again as bootstrap time series. Blocks in this procedure may overlap with each other. This algorithm is equivalent to the subsampling method \cite{Politis1994sub}, a variant of the resampling methods applicable to stationary time series and a particular case of the bootstrap method \cite{Lahiri2003}. Regarding that an optimal block length $\ell^\text{(block)}_\text{opt}$ for the blocking method is also $\ell^\text{(block)}_\text{opt} \sim n^{1/3}$ \cite{Wolff2004}, we find the subsampling method includes the blocking as a special case: If we take only non-overlapping blocks with length $\ell^\text{(block)}_\text{opt}$ from the time series in the subsampling algorithm, that is the optimal blocking. From this viewpoint, the blocking, as well as the stationary bootstrap, belongs to the family of resampling methods for stationary time series.

In summary, the stationary bootstrap method gives error estimation and statistical properties from only one single time series, using a practical algorithm for finding an optimal value of the parameter $p_\text{opt}$. As well as the original bootstrap method \cite{Efron1979,Efron1994}, the stationary bootstrap makes no assumption on the distribution and allows us to access the full (approximate) distribution of desired quantities. Thus, more advanced statistical analyses, such as accelerated estimation of confidence intervals \cite{Efron1987,DiCiccio1996,Efron1994} and statistical hypothesis testing, may be combined with the stationary bootstrap method. The method further provides a systematic way for estimating statistical properties of more complicated quantities, e.g. correlation functions and correlation lengths derived from them, and also the latent heat at a first-order transition that is usually calculated from the distribution function. A simple Python implementation can be found at \url{https://github.com/YoshihikoNishikawa/StationaryBootstrap}.

\section{Numerical test}
\label{sec:num}
In this section, we apply the methods we have reviewed in the previous sections to the actual time series obtained by Monte Carlo simulations. We calculate the squared standard error times the total number of samples, $Mn \times \sigma^2_n$, and the cover rate of the estimated confidence interval as functions of the length of time series $n$. 
The former quantity $Mn \times \sigma^2_n$ should converge to a constant value in the limit $n\rightarrow\infty$.
By regarding the estimation using $10^4$ independent samples as exact, we evaluate the performance of the other methods and the estimation with fewer independent runs.

\subsection{Model}
We consider the two-dimensional ferromagnetic Ising model on a square lattice at the critical temperature. The Hamiltonian of the system is
\begin{equation}
    H\bigpar{\vec s} = -J \sum_{\langle i,j\rangle} s_i s_j,
\end{equation}
with $J > 0$ and $s_i \in \{\pm 1\}$. The summation runs over the pairs of neighboring lattice sites. The linear dimension of the lattice $L = 32$ (the total number of spins $N = L^2$), and the boundary conditions in both directions are periodic. Starting from a random configuration in the high-temperature limit, we equilibrate the system using the single-cluster Wolff algorithm \cite{Wolff1989} with $10^4$ cluster flips. We then use the Metropolis algorithm with sequential update to measure the magnetization $m$ every $10^2$ Monte Carlo sweep. For each time series $\{m_i\}_{i=1,\cdots,n}$, we calculate three commonly studied quantities: the simple average of the magnetization $\overline{m}= \sum_i m_i / n$, the magnetic susceptibility $\chi / N = \overline{m^2} - \overline{m}^2$, and the Binder cumulant $U = \overline{\bigpar{m - \overline{m}}^4} / \overline{\bigpar{m - \overline{m}}}^2$.
The correlation time $\tau_m$ of the dynamics, defined as the time scale of the asymptotic exponential decay of the autocorrelation function of $m$, is $1.2\times 10^3$ sweeps per spin for our setup. Since we measure the magnetization once every $10^2$ Monte Carlo sweeps, $n_m = \tau_m / 10^2 = 1.2 \times 10^1$ is the time-series length corresponding to the correlation time. We rescale the time-series length $n$ by $n_m$ in the following.

As mentioned above, the initial relaxation to the stationary state needs additional care. Using the Wolff cluster algorithm, which virtually eliminates the difficulty at the critical temperature, we consider the ideal case where the time series are sampled from the stationary state in order to study purely the performance of the error estimation technique.

\subsection{Estimation of confidence interval}
To estimate the correct statistical properties of the sampling distribution, we prepare $10^4$ independent time series and calculate the standard errors for $m$, $\chi/N$, and $U$ using the entire data set naively. As a more realistic situation, we take a small number of independent samples, $M = 5$ and $10$, out of $10^4$ time series and calculate the standard errors as well. Using $10^4/M$ standard errors each estimated with $M$ samples, we show the median value of the distribution of the standard error in the following \footnote{We show the median because the empirical distribution of the confidence interval, built with $10^4/M$ blocks, is highly asymmetric with a long tail. In this case the median serves as the best value for a typically obtained result.}.

For the stationary bootstrap method, we estimate $p_\text{opt}$ for each time series \footnote{We should note that, while we have estimated $p_\text{opt}$ using the magnetization and used it for other quantities for simplicity, it is possible to estimate $p_\text{opt}$ for each quantity of interest \cite{Lahiri1999,Lahiri2003}, which would further enhance the performance.}, and calculate the standard error and the cover rate of the estimated $68\%$ confidence interval using the bootstrap percentile method, which we introduced in \Sec{sec:iidboostrap}. We take $10^3$ independent time series to average the standard error and confidence interval. We find, with the stationary bootstrap method, the mean and median values have only a small difference, and thus we show the mean values.

We also apply the blocking method to the magnetization of each time series. In practice, we need to find a plateau in the standard error $\sigma^{\text{(block)}}(R)$ as a function of the number of blocking procedure $R$ for each time series, which corresponds to a fixed point of the renormalization group \cite{Flyvbjerg1989}. However, it is hard to exactly define where the curve enters a plateau region due to the noise, and in many cases no true plateaus even exist. Although an algorithm for automatically selecting the optimal number of blocking was proposed in Ref.~\cite{Jonsson2018}, we take the value at which the absolute derivative of the standard error as a function of the number of blocking
\begin{equation}
    \left|\Delta\sigma^{\text{(block)}}(k)\right| 
    = \left|\frac{\sigma^{\text{(block)}}(k + 1) - \sigma^{\text{(block)}}(k - 1)}2\right|
\end{equation}
becomes the minimum for each time series, and average it over $10^3$ samples for simplicity. As we will show below, this simple procedure works sufficiently well and the estimation becomes precise when $n$ becomes large.

\section{Results}
\label{sec:results}

\subsection{Standard error of $m$, $\chi/N$, and $U$}

\begin{figure}[t]
\includegraphics[width=\linewidth]{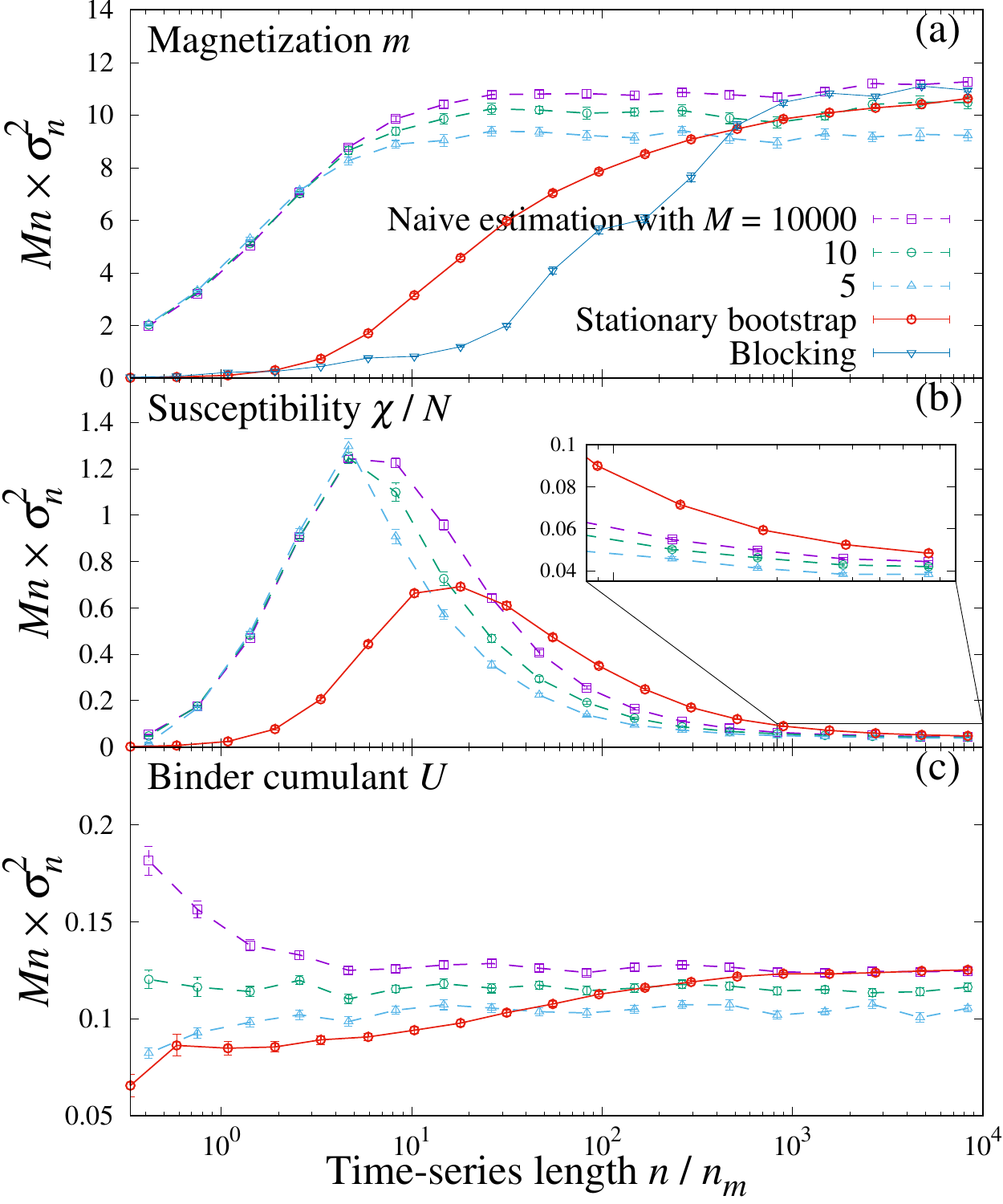}
\caption{The coefficient of the standard error, $Mn \times \sigma_n^2$, as
a function of $n$ for (a) the magnetization $m$, (b) susceptibility $\chi / N$, and (c) Binder cumulant $U$, estimated using $M=10^4$, $10$, $5$ independent runs, the stationary bootstrap, and the blocking method (only for $m$). Inset of (b) shows an enlarged view. The error bars represent the standard error calculated with Efron's bootstrap method.}
\label{fig:standard_error}
\end{figure}

\begin{figure}[t]
\includegraphics[width=\linewidth]{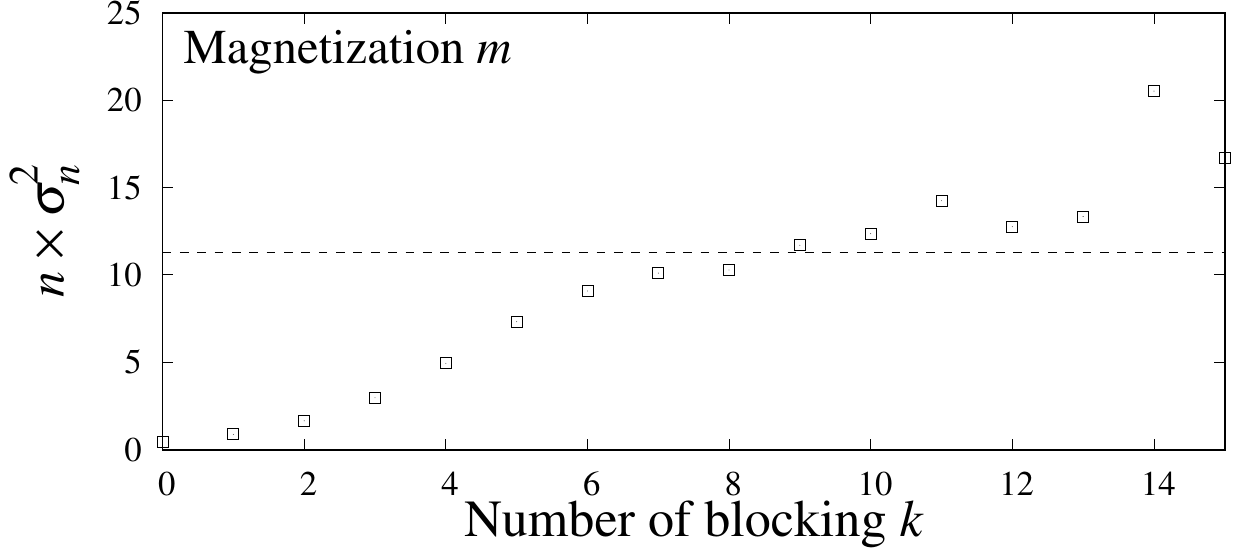}
\caption{A typical curve of $n\times\sigma_n^2$ of the magnetization $m$ obtained by the blocking method. The time-series length $n=10^5$. The dashed line indicates the value of $n\times \sigma_n^2$ estimated using $M=10^4$ independent runs. }
\label{fig:typical_blocking}
\end{figure}

We first discuss the coefficient of the standard error, that is, the squared standard error multiplied by the total number of samples $Mn\times\sigma^2_n$. We compare the values estimated by different methods; the stationary bootstrap, blocking, and simple calculation using $M$ independent runs (see \Fig{fig:standard_error}). At any $n$, the standard error estimated with $M = 10^4$ independent runs should be the most precise. The performances of other methods are evaluated by how much their estimated coefficients are close to the value of $M = 10^4$ independent runs at each $n$.
Since the standard error decays as $(Mn)^{-1/2}$ asymptotically, the coefficient converges to a positive constant value towards $n\to\infty$.

When the number of independent runs $M = 10^4$, $Mn\times\sigma^2_n$ converges at $n / n_m \gtrsim 10^1$ for the magnetization and Binder cumulant, and at $n / n_m \gtrsim 10^4$ for the susceptibility, to the value at $n\to\infty$. 
Recall that $n_m$ corresponds to the correlation time measured in the sample index.
The simple estimation with practical $M = 5, 10$ gives smaller $Mn\times\sigma^2_n$ at large $n$ than that with $M=10^4$. A longer simulation seems not to resolve this problem, as $Mn\times\sigma^2_n$ remains smaller and almost independent of $n$ even when $n / n_m$ reaches $10^4$. 
This may be somewhat surprising regarding how common it is to adopt this method in practical situations. 
We conclude that the simple estimation with a small $M$ does not capture the statistical properties of the distribution correctly, and is unreliable. 
Since we never know the true value of $Mn \times \sigma^2_n$ and its dependence on $M$ in general, we must be cautious of simply using the standard error as the confidence interval unless $M\gg 1$.
We should note that the original bootstrap would be useful to estimate the standard error more precisely for the case with small $M$. Also, bootstrap acceleration methods \cite{Efron1987,DiCiccio1996,Efron1994} would give a better estimation of the $68\%$ confidence interval for the small-$M$ cases.

The stationary bootstrap, which uses only $M=1$ time series for estimating statistical properties, gives $Mn \times \sigma^2_n$ that deviates from the simple estimation with $M=10^4$ runs when the time series is short. Towards larger $n$, it shows a converging behavior for any quantity, and eventually becomes very close to the estimation of $10^4$ samples at large $n$, see \Fig{fig:standard_error}. 
For the magnetization, the blocking method also successfully estimates $n \sigma^2_n$ on average at large $n$ (see \Fig{fig:standard_error}~(a)). However, even when the blocking method works correctly at $n / n_m \simeq 10^4$, it gives a rather noisy curve for each time series and some do not even have a visible plateau; see \Fig{fig:typical_blocking} for a typical curve obtained by the blocking method. This suggests that estimating the standard error for a given time series with the blocking method could be difficult.

It is worthwhile to consider the necessary number of samples in relation to the correlation time of the dynamics $\tau_m \simeq 1.2 \times 10^3$ Monte Carlo sweeps per spin for the magnetization $m$.
The time scale to reach the limiting coefficient $Mn \times \sigma_n^2$ is much longer than $n_m = \tau_m / 10^2$ for all methods. 
This large discrepancy is natural because approximating a distribution with precision needs many samples even in the i.i.d. case.
Yet, it is worth noting that the total number of samples needed to reach the limiting $Mn \times \sigma_n^2$ value varies with methods: For the simple calculation using $M=10^4$ independent runs, the number is about $4\times 10^2 \times M \simeq 4\times 10^6$, while that for the stationary bootstrap and the blocking method is about $10^5$, which is significantly smaller than the simple calculation using $M=10^4$ independent runs.

In the next subsection, we directly test whether each method works properly and gives a reasonable estimate for the confidence interval by calculating the cover rate.

\subsection{Cover rate of estimated confidence intervals}
\label{subsec:cover}

\begin{figure}[t]
\includegraphics[width=\linewidth]{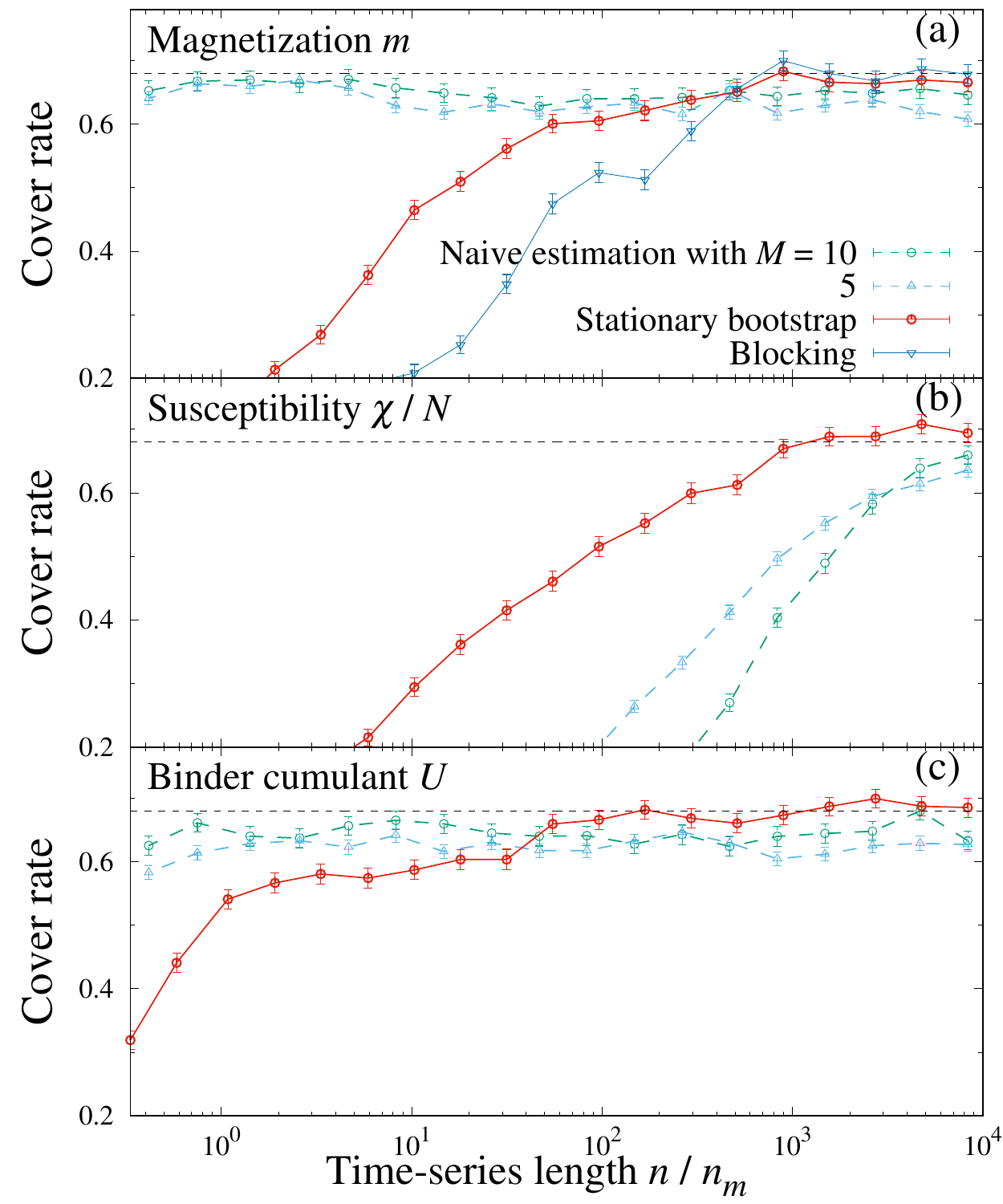}%
\caption{The cover rate of estimated $68\%$ confidence intervals for (a) the magnetization $m$, (b) susceptibility $\chi / N$, and (c) Binder cumulant $U$, estimated using $M=10$, $5$ independent runs, the stationary bootstrap, and the blocking method (only for $m$). The dashed line indicates $68\%$. The error bars represent the standard error calculated with Efron's bootstrap method.}
\label{fig:cover_rate}
\end{figure}

In the previous section, we compared the speed of convergence in terms of the standard error. 
A more direct way to compare the different methods is to see the ``cover rate" of the estimated confidence intervals. Although this is not a quantity usually measured for practical purposes, it gives a reasonable estimate of how reliable the different methods are. 

We use the bootstrap percentile method \cite{Efron1994,DiCiccio1996} introduced in \Sec{sec:iidboostrap}. This simple method is applicable to any distribution and allows us to calculate a confidence interval. 
Whether or not the estimated confidence interval works properly can be assessed by calculating the cover rate, that is, the ratio of independent samples having the true value within the confidence intervals: If the estimated confidence interval is accurate, the cover rate is equal to $68\%$.

Here, with the stationary bootstrap method, we estimate the $68\%$ confidence interval. The cover rate is then calculated using $10^3$ independent runs. Since the true values are unknown apart from the simple mean, which is $0$, we use the simple averages over $10^4$ independent runs as the true values.
For comparison, we also calculate the cover rate of the standard error estimated using $M = 5$ and $10$ independent runs and that of the blocking method. For the standard error of $M$ independent runs, the cover rate is calculated as the ratio of $10^4/M$ sets of runs having the true value in the range of the $68\%$ confidence interval. We use $[\bar{A}_n -\sigma_n, \bar{A}_n + \sigma_n]$ as the $68\%$ confidence interval for these cases.

\Fig{fig:cover_rate} shows the cover rate for the magnetization $m$, susceptibility $\chi/N$, and the Binder cumulant $U$. When $M=5$ and $10$, the cover rate has a clear systematic deviation from $68\%$ and does not reach the correct value even at $n / n_m = 10^4$. The deviation of the cover rate when $M=5$ and $10$ is consistent with the coefficient of the standard error, $Mn\times \sigma_n^2$, that is also lower than the correct value, see \Fig{fig:standard_error}. This indicates that the estimated statistical values are over-confident when $M$ is small. Again, the standard error simply calculated using small $M$ is thus unreliable and should not be used in general, while, as noted in the previous section, the original bootstrap would resolve the issue of this method. The cover rate of the stationary bootstrap method, on the other hand, reaches a value very close to $68\%$ at large $n$, meaning that the estimated confidence interval indeed works correctly for all the physical quantities. The cover rate of the blocking method for $m$ also reaches $68\%$ at large $n$, but more slowly compared to the stationary bootstrap method.

Apparently, the confidence interval for the susceptibility $\chi / N$ is the most difficult to estimate precisely, as it is the slowest to reach $68\%$, see \Fig{fig:cover_rate}~(b). The difficulty comes from the strong asymmetric shape of the sampling distribution of $\chi / N$ for small $n$, as shown in \Fig{fig:NonGaussianIsBad}~(a). Other quantities involving fluctuations are expected to have the same difficulty. This suggests that, for those having asymmetric distributions, more sophisticated acceleration methods \cite{Efron1987,DiCiccio1996,Efron1994} would give a much better estimate for the confidence interval. It is an interesting question whether and how much acceleration methods improve the accuracy \cite{Hall1988} for time-correlated samples, as in the case of i.i.d. samples.

\section{Summary and discussion}
\label{sec:summary}

In summary, we reviewed some methods for estimating statistical properties of actual MCMC-generated samples that have temporal correlations, with particular focus on the stationary bootstrap method. 
We then discussed the connection between the family of resampling methods (including the stationary bootstrap) and blocking method, and argued the blocking method can be regarded as a special case of the resampling methods.
Using a realistic bench-marking data set, we critically assessed the performance of the statistical methods. The simple way of estimating the standard error using several independent runs was revealed to be erroneous and may be over-confident when the number of independent runs $M$ is too small. The stationary bootstrap method, which uses only one time series, are much more precise than the simple way with a small $M$. 

When $M$ independent time series are readily available, assembling an $M$-times-longer time series from them and then applying the stationary bootstrap method to it should be more precise than the simple estimation using several independent runs.
The combined time series could be regarded as a MCMC data with occasional insertion of a perfect Monte Carlo update where a configuration is drawn with the equilibrium probability. The violation of the temporal stationarity is minimal in this case, and so should be the effect to the accuracy of the stationary bootstrap.

One of the major advantages of the stationary bootstrap is its ability to access the full sampling distribution and applicability to a large class of quantities, such as the specific heat, the Binder cumulant, normalized spatial correlations, and even the latent heat at a first-order transition. More complicated quantities involving multiple variables, such as the second-moment correlation length \cite{Cooper1982,Campostrini2001,Hasenbusch2005,Nishikawa2016} and the bulk and shear moduli \cite{Xu2012}, are within the scope of the method as well \cite{Politis1994}. Another potential use of the method is to reliably estimate the peak locations of measured functions such as the structure factor.

\begin{figure}[t]
    \includegraphics[width=\linewidth]{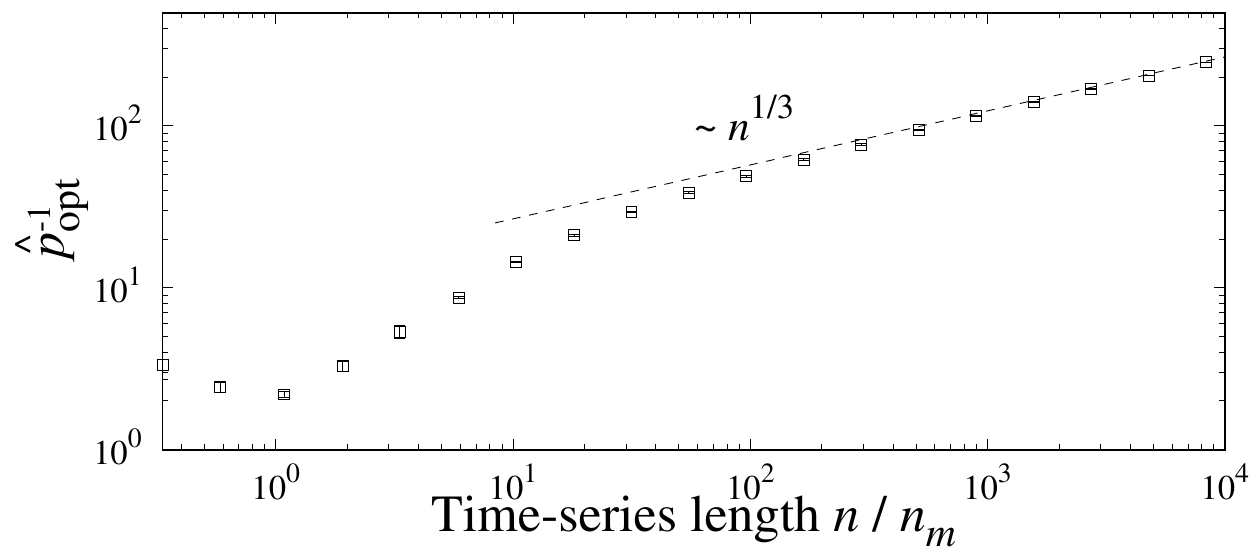}
    \caption{The inverse of the estimated probability $\hat p^{-1}_\text{opt}$ as a function of $n / n_m$. At $n / n_m \gtrsim 10^3$, $\hat p^{-1}_\text{opt}$ asymptotically grows as $\sim n^{1/3}$ as expected from \eq{eq:popt}.}
    \label{fig:inv_popt}
\end{figure}

Finally, it is crucial to have enough samples in the data compared to the correlation time. In the stationary bootstrap method, this could be checked by comparing inverse of the estimated optimal probability $\hat p^{-1}_\text{opt}$ with the time-series length $n$: Since $p^{-1}\sim n^{1/3}$ is a necessary condition for the convergence of the stationary bootstrap \cite{Politis1994}, we should confirm if $\hat p^{-1}_\text{opt}$ actually grows as $\sim n^{1/3}$. In our numerical test, $\hat p^{-1}_\text{opt}$ indeed becomes asymptotically $\sim n^{1/3}$ at $n / n_m \gtrsim 10^3$, see \Fig{fig:inv_popt}. Another valid sanity check would be to see if the estimated error scales as $n^{-1/2}$. Although some advanced acceleration methods \cite{Efron1987,Efron1994,DiCiccio1996} would be useful for better estimation of errors and confidence intervals when the time series is rather short, the best is to simply perform a much longer simulation and obtain a long enough time series. Otherwise we get incorrect statistical errors with any method and would reach a wrong scientific conclusion.

\begin{acknowledgments}
The authors would like to thank W.~Krauth and K.~Hukushima for their critical reading of the manuscript and useful comments. They also thank Y.~Ozeki and A.~W.~Sandvik for discussions. 
JT was supported by the National Science Foundation (NSF) Focused Research Hub for Theoretical Physics, Center for Quantum Information and Control (CQuIC) under Grant No. PHY-2116246 and supported by the U.S. Department of Energy (DOE), Office of Science, National Quantum Information Science Research Centers, Quantum Systems Accelerator (QSA).
\end{acknowledgments}

\appendix

\section{Performance of Efron's original bootstrap method when applied to time series}
\label{app:original_bootstrap}

\begin{figure}[t]
\includegraphics[width=\linewidth]{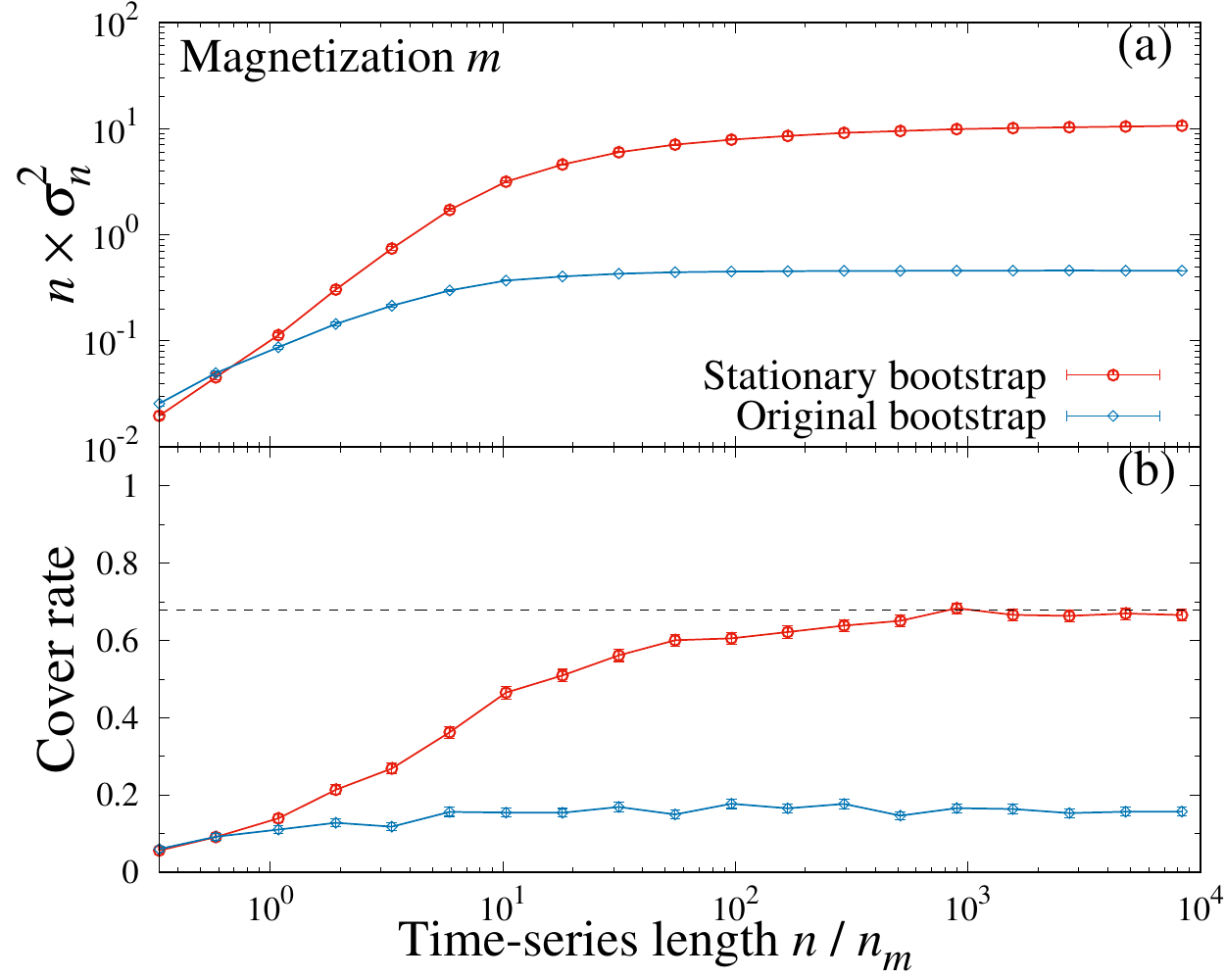}%
\caption{(a) The coefficients of the standard error $n\times \sigma_n^2$ and (b) the cover rates of estimated $68\%$ confidence intervals for the magnetization $m$ estimated by the the stationary bootstrap and the original bootstrap method. The dashed line in (b) indicates $68\%$. The error bars represent the standard error calculated with Efron's bootstrap method.}
\label{fig:bs+sbs}
\end{figure}

In this appendix, we study and discuss the performance of Efron's original bootstrap when applied to time-correlated samples. We calculate the coefficient of the standard error $n\times\sigma_n^2$ and the cover rate of estimated $68\%$ confidence intervals for the magnetization $m$ with the bootstrap percentile method. \Fig{fig:bs+sbs} shows them as functions of the time-series length $n$, together with those of the stationary bootstrap method. The original bootstrap method clearly fails to estimate the standard error: The coefficient of the standard error $n\times \sigma_n^2$ differs from that of the stationary bootstrap method by an order of magnitude, and the cover rate is far below $68\%$. When applied to time series, the original bootstrap, as designed for i.i.d. samples, randomizes the order of time-correlated samples. Hence, resultant bootstrap samples lose the time correlation encoded in the original samples. The argument made in Ref.~\cite{Newman1999}, saying the original bootstrap method gives a reasonable error even if applied to time-correlated samples, is thus incorrect. If one still wishes to use Efron's bootstrap method to a time series, the time interval between two successive samples needs to be potentially very long, such as $2\tau$ with $\tau$ the correlation time. This strategy makes the number of samples in the time series fewer, possibly leading to worse statistics.

\section{Optimal parameter $p_\text{opt}$}
\label{app:popt}

In this appendix, a (semi)automatic algorithm to find the optimal value $p_\text{opt}$ for a time series is presented. Following Ref.~\cite{Politis2004}, we calculate the unnormalized autocorrelation function \eq{eq:unautocorr} from the time series, and numerically estimate \eq{eq:G} and \eq{eq:D}.

Using the unnormalized autocorrelation function
\begin{equation}
\hat C(k) = \sum_{i = 1}^{N - |k|} (x_i - \overline{x})(x_{i + |k|} - \overline{x}) / N
\end{equation}
with $\overline x = \sum_i x_i / n$, we find $T = 2\hat t$ with $\hat t$ the smallest number satisfying 
\begin{equation}
|\hat C(\hat t + k) / \hat C(0)| < c \sqrt{\log_{10} N / N}
\end{equation}
for $k = 1, 2, \cdots, K_n$ with a constant $c > 0$ and $K_n = o(\log_{10}n)$. The constant $c$ and $K_n$ should be carefully chosen not to underestimate $T$ although any choice of $c > 0$ and $0 < K_n < n$ should work asymptotically \cite{Politis2003}. $c=2$ and $K_n = \max\bigpar{5, \sqrt{\log_{10}n}}$ are suggested as a safe choice in Refs.~\cite{Politis2003,Politis2004}. Since $\hat C(k)$ is calculated only from one single time series, it is noisy especially at long time. By finding $T$ in this manner, we estimate a rough upper bound for the correlation time from the rather noisy autocorrelation. Note that, when $\hat t$ and $T$ are comparable to $n$, the time series is clearly not long enough and we need much more samples for precise estimation. With $T$ and a window function \cite{Politis1995,Politis2003}
\begin{equation}
w(k) = \left\{
\begin{array}{lll}
1 & \text{ if } |k| \in [0, 1/2], \\
2(1 - |k|) & \text{ if } |k| \in [1/2, 1],\\
0 & \text{otherwise},
\end{array}\right.
\end{equation}
we calculate 
\begin{align}
&\hat G = 2\sum_{k=0}^{T}w(k/T)|k|\hat C(k), \\
&\hat D = 2 \left( \hat C(0) + 2\sum_{k=1}^{T}w(k/T)\hat C(k) \right)^2,
\end{align}
and finally reach
\begin{equation}
\hat p_\text{opt} = \left( \frac{2\hat G^2}{\hat D}\right)^{-1/3}n^{-1/3},
\end{equation}
which is an estimation of the optimal parameter $p_\text{opt}$.

If an estimated $\hat p_\text{opt}$ is larger than $1$, the sample correlation in the time series is negligible and the block size in the stationary bootstrap method is $1$, meaning that the method is equivalent to the original bootstrap method for i.i.d. samples in this case.

\bibliography{refs}

\end{document}